\documentclass{Interspeech}
\usepackage{amsmath}
\usepackage{hyperref}

\interspeechcameraready

\title{CabinSep: IR-Augmented Mask-Based MVDR for Real-Time In-Car Speech Separation with Distributed Heterogeneous Arrays}

\author[affiliation={1}]{Runduo}{Han}
\author[affiliation={2}]{Yanxin}{Hu}
\author[affiliation={1}]{Yihui}{Fu}
\author[affiliation={1}]{Zihan}{Zhang}
\author[affiliation={1}]{Yukai}{Jv}
\author[affiliation={2}]{Li}{Chen}
\author[affiliation={1^{*}}]{Lei}{Xie}

\affiliation{Audio, Speech and Language Processing Group (ASLP@NPU), School of Computer Science}{Northwestern Polytechnical University}{Xi'an, China}
\affiliation{}{ Shanghai ZEEKR Blue New Energy Technology Co., Ltd.}{China}
\email{
  \{rdhan, zhzhang, nwpu2016303311\} @mail.nwpu.edu.cn,
  arrowhyx@foxmail.com, 
  felixfuyihui@gmail.com, 
  lichenbook1@163.com, 
  lxie@nwpu.edu.cn\thanks{* Corresponding author.}
}


\begin{document}
\maketitle
\begin{abstract}
\vspace{-2pt}

Separating overlapping speech from multiple speakers is crucial for effective human-vehicle interaction. This paper proposes \texttt{CabinSep}, a lightweight neural mask-based minimum variance distortionless response (MVDR) speech separation approach, to reduce speech recognition errors in back-end automatic speech recognition (ASR) models. Our contributions are threefold: First, we utilize channel information to extract spatial features, which improves the estimation of speech and noise masks. Second, we employ MVDR during inference, reducing speech distortion to make it more ASR-friendly. Third, we introduce a data augmentation method combining simulated and real-recorded impulse responses (IRs), improving speaker localization at zone boundaries and further reducing speech recognition errors. With a computational complexity of only 0.4 GMACs, \texttt{CabinSep} achieves a 17.5\% relative reduction in speech recognition error rate in a real-recorded dataset compared to the state-of-the-art DualSep model\footnote{Demos are available at: https://cabinsep.github.io/cabinsep/}. 

\end{abstract}

\vspace{-4pt}
\section{Introduction}
\vspace{-2pt}


Speech interaction is crucial for in-car intelligence, with automatic speech recognition (ASR) as a key gateway for human-vehicle interaction. Speech recognition accuracy directly impacts interaction efficiency and user experience~\cite{H-V,H-V2,ICC,Conversation,ASR}. However, when multiple passengers interact with the car simultaneously, overlapping speech becomes inevitable, posing significant challenges to the ASR system~\cite{CHiME3,AV_ASR,NPU-MISP2022,mu2024,neural_fcasa}. Therefore, it is necessary to apply speech separation systems to separate overlapping speech and distinguish speakers before delivering speech recognition~\cite{ROSM}.

\begin{figure*}[!htbp]
\vspace{-40pt}
  \centering
  \includegraphics[height=0.24\textheight, width=1\linewidth]{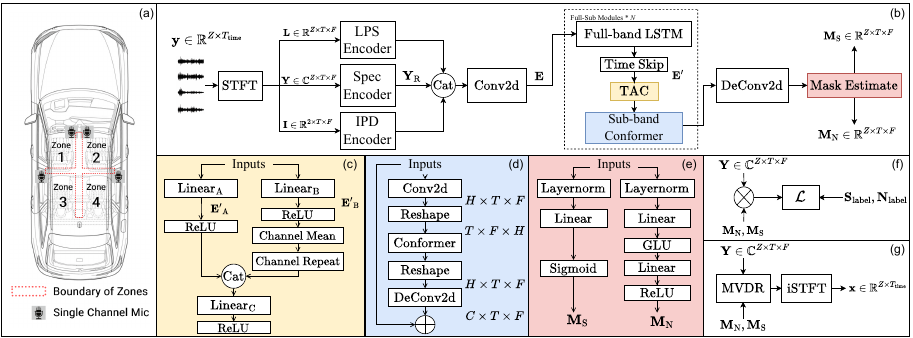}
  \vspace{-16pt}
  \caption{{The overall structure of \texttt{CabinSep}. (a)~in-car speech separation scenario; (b)~model architecture; (c)~transform-average-concatenate (TAC) module; (d)~sub-band-conformer module; (e)~mask estimate module; (f)~Training procedure; (g)~Inference procedure; $\mathbf{y}$ represents $Z$ channel audio mixture; $\mathbf{Y}$ represents complex spectrum of $\mathbf{y}$; $\mathbf{L}$ represents $\mathbf{LPS}$; $\mathbf{I}$ represents $\mathbf{IPD}$; $\mathbf{M}_\mathrm{S}$,$\mathbf{M}_\mathrm{N}$ represent estimated speech and noise masks; $\mathbf{S}_\mathrm{label}$, $\mathbf{N}_\mathrm{label}$ represent speech and noise labels; $\mathbf{x}$ represents separated clean speech; $\mathcal{L}$ represents the loss function.}}
  \label{fig:sturcture}
\vspace{-16pt}
\end{figure*}

The neural mask-based minimum variance distortionless response (MVDR) speech separation scheme has been proven effective in improving speech recognition performance by ensuring undistorted speech~\cite{mmvdr2016, imporved, joint}. However, during the joint training of the neural network and MVDR, the matrix inversion operation in MVDR causes numerical instability, leading to unstable model convergence~\cite{fast}. Additionally, noise elimination of this scheme is unsatisfactory because MVDR weight selection does not achieve optimal noise suppression~\cite{ADL-MVDR}. Recent works~\cite{ADL-MVDR,Multi,embB,fasnet-tac, dualsep} adopt speech separation schemes that directly use the neural network's output as separated speech, of which~\cite{dualsep, zoneformer} approaches achieve remarkable performance in intrusive speech quality metrics including scale-invariant signal-to-noise ratio (SI-SNR)\cite{sisnr}. However, due to the inherent problem of neural networks, nonlinear distortion is introduced in complicated real-world scenarios, negatively affecting ASR performance. Although these advanced models~\cite{zoneformer, dualsep} have a relatively small number of parameters, their computational complexity exceeds 1 GMACs, making them challenging to deploy in cars practically. Furthermore, the irregular structures of car cabins make it difficult to simulate training data that matches real-world scenarios. Meanwhile, back-end systems sometimes need to respond based on the speaker's position in in-car speech interactions. As shown in Figure~\ref{fig:sturcture}(a), accurately locating passengers is challenging when they speak at the boundary of a zone, which previous studies have not adequately addressed.

We propose \texttt{CabinSep}, a plug-and-play streaming speech separation approach for real-world in-car scenarios that improves ASR performance to a great extend. To address nonlinear speech distortion in ASR systems~\cite{MISP2023,single}, we adopt MVDR during inference to avoid numerical instability and reduce distortion. Furthermore, we introduce a dual-mask estimation mechanism where an auxiliary noise mask complements the speech mask, improving MVDR's noise suppression capabilities. Effectively utilizing spatial information in feature channels aids speech separation tasks, such as the transform-average-concatenate (TAC) module~\cite{fasnet-tac,neural_fcasa}. However, it involves considerable computational complexity. We propose a time skip cascaded TAC module that processes only half of the time frames, reducing computational complexity significantly while maintaining performance. Unlike previous solutions~\cite{ADL-MVDR, zoneformer}, our proposed \texttt{CabinSep} skips the highly accurate direction of arrival (DOA) estimation, which allows for a simpler heterogeneous microphone array design, with each zone in the cabin corresponding to a single-channel microphone, thereby reducing production costs.


To simulate training data that better matches in-car communication scenarios, we propose a data augmentation method that combines simulated impulse responses (IRs) with real-recorded IRs~\cite{IRs}. Notably, this method includes an effective IRs mixing strategy where real-recorded IRs are used for the microphone of the speaker's zone, while simulated IRs are used for other zones. This approach enables \texttt{CabinSep} to accurately localize the zone of the speaker even when the speaker sits at the zone's boundary.

Our proposed \texttt{CabinSep} approach is evaluated by speech recognition accuracy on two open-source pre-trained ASR models, WeNet~\cite{wenet} and SenseVoice~\cite{sensevoice}, which validates the robustness of \texttt{CabinSep} approach across different back-end ASR models architectures. All test data are real-recorded audio from an electric vehicle to better match real-world scenarios. 
This includes data recorded while the car is stationary or during motion. When the vehicle is moving, the speech is disturbed by complicated background noise, including wind, wheel, engine, etc~\cite{liauto}.
Experimental results show that our proposed \texttt{CabinSep} achieves a 17.5\% relative reduction in speech recognition error rate compared to the state-of-the-art (SOTA) approach DualSep~\cite{dualsep} for in-car speech separation and recognition, with a computational complexity of only 0.4 GMACs and 0.21 real-time factor (RTF) using the single-core of Qualcomm SA8295P in-car CPU. 


\vspace{-4pt}
\section{Problem Formulation}
As shown in Figure~\ref{fig:sturcture}(a), we focus on in-car speech separation. We divide the car cabin into $Z$ zones, with each zone corresponding to a single channel microphone, and at most one person speaks in each zone. Suppose there are $P$ person speaking in the car ($P\leq Z$), the clean speech corresponding to each zone is set as $\mathbf{s(z)}$, and the signal recorded by each microphone is $\mathbf{{y}_{i}}$, which can be expressed as:
\vspace{-4pt}
\begin{equation}
\mathbf{x}_{i}(z) = \mathbf{r}_{i}{(z)} * \mathbf{s}(z),
\end{equation}
\vspace{-16pt}
\begin{equation}
\mathbf{y}_{i} = \sum_{z=1}^{Z} \mathbf{x}_{i}(z) + \mathbf{v}_i ,
\end{equation}
\vspace{-1pt}
where $z$ represents the zone index in the cabin,  $i \in \{1,...,Z\}$ represents the microphone index, $\mathbf{r}$ represents the IRs, and $\mathbf{v}$ is the background noise recieved by micriphone. We aim to estimate the separated clean speech $\mathbf{x}_z(z)$ corresponding to each zone.

Due to the varying postures of passengers, the position of the speech source location within each zone can be arbitrary. When a speaker adopts a relatively standard sitting posture, it becomes easier to accurately separate each speaker's voice into the correct zone, which is called ``standard postures". However, when a speaker's sitting posture is non-standard, their speech may occur at the boundary between adjacent zones, as indicated by the area within the red dotted line in Figure~\ref{fig:sturcture}(a), which is called ``non-standard postures". This may cause the speech separation system to incorrectly localize speech to adjacent zones.

\vspace{-4pt}
\section{Method}
\vspace{-4pt}
\subsection{System overview}
\vspace{-4pt}
The overall architecture of our proposed \texttt{CabinSep} is shown in Figure~\ref{fig:sturcture}(b). First, the $Z$-channel audio mixture $\mathbf{y}$ is transformed by the Short-Time Fourier Transform (STFT) to obtain the T-F spectrum $\mathbf{Y} \in \mathbb{C}^{Z \times T \times F}$, where $Z$ represents the channel dimension, $T$ represents the time dimension, and $F$ represents the frequency dimension. The log power spectrum (LPS) $\mathbf{L}$ and interaural phase difference (IPD) $\mathbf{I}$~\cite{zoneformer,ipd} are then derived from $\mathbf{Y}$, which contain abundant spectral and spatial information, respectively. Next, we adopt three encoders to embed $\mathbf{Y}$, $\mathbf{L}$, and $\mathbf{I}$, respectively. These embeddings are concatenated along the channel dimension and passed through a convolution layer, resulting in $\mathbf{E} \in \mathbb{R}^{C \times T \times F}$. Multiple full-sub modules are then adopted to refine these T-F embeddings~\cite{fsbnet}, and the output is restored to $Z \times T \times F$ through deconvolution. Finally, the mask estimate module is adopted to simultaneously estimate speech mask $\mathbf{M}_\mathrm{S}$ and noise mask $\mathbf{M}_\mathrm{N}$.
During inference, as shown in Figure~\ref{fig:sturcture}(g), a streaming MVDR~\cite{mvdr} utilizes the $\mathbf{M}_\mathrm{S}$ and $\mathbf{M}_\mathrm{N}$ along with $\mathbf{Y}$ as input to obtain the separated clean speech $\mathbf{X}_i$ for each zone:
\vspace{-4pt}
\begin{equation}
\mathbf{X}_{i,t,f} = \mathbf{W}_{i,t,f}^\mathrm{H}\mathbf{Y}_{t,f},
\end{equation}
\vspace{-12pt}
\begin{equation}
\mathbf{W}_{i,t,f} = \frac{\Psi^{-1}_{i,t,f} \Phi_{i,t,f} \mathbf{e}}{\operatorname{tr}(\Psi^{-1}_{i,t,f} \Phi_{i,t,f})}
\end{equation}
where $\mathbf{W_{i,t,f}}$ is the MVDR coefficient vector for each zone, and $\mathbf{e}$ is a one-hot vector with $1$ at the reference microphone position. The matrices $\Psi_{i}$ and $\Phi_{i}$ represent the spatial covariance matrices of the target speech and overlapping interference, derived from $\mathbf{M}_\mathrm{S}$ and $\mathbf{M}_\mathrm{N}$. The detailed calculation method is described in~\cite{ROSM,mvdr}.

\vspace{-4pt}
\subsection{Spec Encoder, LPS Encoder and IPD Encoder}
\vspace{-4pt}
The proposed \texttt{CabinSep} consists of three encoders. Each encoder consists of two convolution layers, and the ReLU is used as the activation after each convolution layer. The real and imaginary parts of $\mathbf{Y}$ are stacked along the channel dimension to obtain $\mathbf{Y}_\mathrm{R}\in \mathbb{R}^{  2Z \times T \times F}$, which is then sent into the Spec encoder. 
The LPS encoder takes the $\mathbf{L}$ as input, which is defined as $\mathbf{L} = \log(|\mathbf{Y}_\mathrm{R}|^2) \in \mathbb{R}^{Z \times T \times F} $.
Due to the audio aliasing caused by large microphone spacing in the back-row (zone 3 and zone 4), the IPD information between these two microphones becomes unusable. Therefore, we only utilize the IPD information $\mathbf{I}$ between the front-row (zone 1 and zone 2) microphones with smaller microphone spacing.
Following the method in~\cite{ipd}, $\mathbf{I}$ is calculated by:
\begin{equation}
\mathbf{I} = \left[ \cos(\theta_{1,2}), \sin(\theta_{1,2}) \right] \in \mathbb{R}^{2 \times T \times F} \text{ ,}
\label{ipd}
\end{equation}
where $\theta_{1,2} = \angle \mathbf{Y}_{1} - \angle \mathbf{Y}_{2}$ represent the phases difference between the signals received by the microphones with index 1 and 2 in zone 1 and zone 2, respectively. 
\vspace{-4pt}
\subsection{Full-Sub Modules}
\vspace{-4pt}
Embedded features are then refined by $N$ full-sub modules. The value of $N$ can be selected according to the limitations of parameters and computational complexity. Each full-sub module comprises three submodules: full-band-LSTM, TAC, and Sub-band-conformer. The full-band-LSTM is used to process the full-band information, and its structure is the same as that in~\cite{fsbnet}.

The structure of the TAC module is shown in Figure~\ref{fig:sturcture}(c). Refering to~\cite{neural_fcasa, fasnet-tac}, TAC models spatial features across different channels. Considering that the computational complexity of the TAC module is relatively large, we propose a time skip operation before TAC. Specifically, we randomly select either the 1st or 2nd frame as the starting time frame, then choose frames to be processed at intervals of 1. The time skip operation halves the number of time frames, and the obtained feature $\mathbf{E}' \in \mathbb{R}^{C \times \frac{T}{2} \times F}$ serves as the input for the TAC. 
Since TAC emphasizes information across channels and is insensitive to time frames, incorporating the time skip operation reduces the computational complexity of TAC by half with little impact on performance.
In TAC, two fully connected layers, $\mathrm{Linear}_\mathrm{A}$ and $\mathrm{Linear}_\mathrm{B}$, first reduce the dimension of the channel dimension of the input feature  to obtain $\mathbf{E}_\mathrm{A}'$ and $\mathbf{E}_\mathrm{B}'$:
\vspace{-4pt}
\begin{equation}
\mathbf{E}' \in \mathbb{R}^{ C\times  \frac{T}{2} \times F} 
\to \mathbf{E}_\mathrm{A}'\in \mathbb{R}^{ \frac{C}{d} \times  \frac{T}{2} \times F}, 
 \mathbf{E}_\mathrm{B}'\in \mathbb{R}^{ \frac{C}{d} \times  \frac{T}{2} \times F}
 \text{ ,}
\end{equation}
where $d$ is the compression ratio of the channel dimension. Then, the ReLU-activated $\mathbf{E}'_\mathrm{B}$ is averaged along the channel dimension and repeatedly stacked. It is then concatenated with the ReLU-activated $\mathbf{E}'_\mathrm{A}$ along the channel dimension to obtain $\mathbf{E}'_\mathrm{cat}$, after which a fully connected layer $\mathrm{Linear}_\mathrm{C}$ is used to restore the dimension.
After TAC, the processed time frames selected by the time skip operation are recombined with the unprocessed time frames.

The sub-band-conformer models the features in sub-bands. Its structure is shown in Figure~\ref{fig:sturcture}(d), and it is generally consistent with that of~\cite{fsbnet}, except that the LSTM is replaced with the conformerr~\cite{conformer, conformer2, uformer, distil-dccrn}. Due to the small number of parameters and low computational complexity associated with the sub-band processing structure, replacing the LSTM with the conformer introduces only a few additional parameters and has minimal impact on computational complexity, while significantly enhancing the temporal modeling ability.

\vspace{-4pt}
\subsection{Masks Estimate and Model Training}
\vspace{-4pt}
The mask estimate module, shown in Figure~\ref{fig:sturcture}(e), estimates both the speech and noise masks using multiple linear layers. During training, the estimated speech and noise masks are applied to the multi-channel audio mixtures to obtain the separated speech and noise, and the losses are calculated with the corresponding labels. The loss function consists of $L_{\textrm{Fbank-MAE}}$ and $L_{\textrm{SI-SNR}}$~\cite{sisnr} where $L_{Fbank-MAE}$ is the mean absolute error (MAE) loss calculated between the fbank feature of outputs and labels~\cite{fbank}. The composition of the overall loss function is:
\vspace{-5pt}
\begin{equation}
\begin{aligned}
\mathcal{L} = & \ \alpha \mathcal{L}_{\text{FBank-MAE}}(\mathbf{S}, \mathbf{S_{label}}) + \beta \mathcal{L}_{\text{SI-SNR}}(\mathbf{S}, \mathbf{S_{label}}) \\
& + \gamma \mathcal{L}_{\text{FBank-MAE}}(\mathbf{N}, \mathbf{N_{label}}) \text{ ,}
\end{aligned}
\end{equation}
where $\mathbf{S}$, $\mathbf{N}$ represent the separated speech and noise, and $\mathbf{S_{label}}$ and $\mathbf{N_{label}}$ are corresponding labels. $\alpha$, $\beta$ and $\gamma$ represent the weight of those loss functions. We set $\alpha = 0.01$, $\beta = 1$ and $\gamma = 0.01$ to balance the magnitude.

We adopt a two-stage training strategy to address inaccurate positioning caused by ``non-standard postures" and further improve the speech recognition accuracy of back-end ASR models. In the first stage, we use simulated IRs for data augmentation. In the second stage, the model is finetuned by the augmented data using a mixture of simulated and real-recorded IRs. 



\vspace{-8pt}
\section{Experiment}
\vspace{-4pt}
\begin{table*}[!htbp]
\vspace{-40pt}
    \centering
    \caption{Comparison of CER on real-recorded speech recognition test set of the 1st-stage model trained with simulated IRs.}
    \vspace{-8pt}
    \renewcommand{\arraystretch}{0.6}
    \resizebox{\textwidth}{!}{
        \begin{tabular}{c l c c c c c c c c c}
            \toprule
            \# & System & Causal & Params M & GMACs & \multicolumn{3}{c}{WeNet~\cite{wenet}} & \multicolumn{3}{c}{SenseVoice~\cite{sensevoice}} \\ 
            \cmidrule(lr){6-8} \cmidrule(lr){9-11}
            & & & & & static CER\% & motion CER\% & average CER\% & static CER\% & motion CER\% & average CER\% \\ 
            \midrule
            1 & Unprocessed Mixture & - & - & - & 62.34 & 42.74 & 55.42 & 39.37 & 20.82 & 32.82 \\ 
            2 & FasNet-TAC & \checkmark & 2.77 & 10.35 & 26.50 & 44.57 & 32.88 & 13.72 & 23.69 & 17.24 \\ 
            3 & DualSep-S & \checkmark & 0.83 & 1.07 & 18.02 & 33.32 & 23.42 & 11.48 & 18.7 & 14.03 \\ 
            4 & DualSep-L & \checkmark & 1.12 & 1.52 & 17.63 & 30.76 & 22.26 & 11.52 & 16.97 & 13.44 \\ 
            5 & CabinSep-S & \checkmark & 1.09 & 0.40 & 16.87 & 22.09 & 18.36 & 10.77 & 13.44 & 11.53 \\ 
            6 & CabinSep-M & \checkmark & 2.24 & 0.62 & 16.46 & \textbf{20.96} & 17.74 & 10.64 & \textbf{13.07} & 11.33 \\ 
            7 & CabinSep-L & \checkmark & 3.43 & 1.22 & \textbf{15.74} & 21.48 & \textbf{17.38} & \textbf{10.53} & 13.28 & \textbf{11.31} \\ 
            \midrule
            7-1 & -MVDR & \checkmark & 3.43 & 1.18 & 26.6 & 42.57 & 31.16 & 14.84 & 23.47 & 17.30 \\ 
            7-2 & +time skip & \checkmark & 3.43 & 0.83 & 16.39 & 21.29 & 17.79 & 10.70 & 13.45 & 11.48 \\ 
            7-3 & -conformer & \checkmark & 3.42 & 1.19 & 16.22 & 22.24 & 17.94 & 10.68 & 13.50 & 11.49 \\ 
            7-4 & -conformer-TAC & \checkmark & 3.02 & 0.43 & 18.08 & 25.42 & 20.18 & 10.85 & 13.98 & 11.74 \\ 
            7-5 & -conformer-TAC-LPS-IPD & \checkmark & 3.02 & 0.38 & 19.00 & 26.48 & 21.13 & 10.98 & 14.84 & 12.08 \\ 
            7-6 & -conformer-TAC-Noise Mask & \checkmark & 2.83 & 0.38 & 19.02 & 26.35 & 21.11 & 10.98 & 14.55 & 12.00 \\ 
            7-7 & +chunk & \checkmark & 3.43 & 1.22 & 15.96 & 21.26 & 17.47 & 10.56 & 13.27 & 11.33 \\
            \bottomrule
        \end{tabular}
    }
    \label{tab:comparison}
\vspace{-16pt}
\end{table*}
\begin{table}[!htbp]
    \centering
    \caption{CER results obtained by WeNet after finetuning the 1st-stage CabinSep-L using real-recorded IRs. $\text{NSPA}$ represents the positioning accuracy rate in ``non-standard posture".}
    \vspace{-8pt}
    \resizebox{\columnwidth}{!}{
    \begin{tabular}{l c c c c c c c c c c}
        \toprule
         Type & \multicolumn{2}{c}{1st-stage IRs} & \multicolumn{2}{c}{mixed real-recorded IRs} & \multicolumn{2}{c}{added real-recorded IRs} & \multicolumn{2}{c}{only real-recorded IRs} \\ 
        \cmidrule(lr){2-3} \cmidrule(lr){4-5} \cmidrule(lr){6-7} \cmidrule(lr){8-9}
        & CER\%$\downarrow$ & NSPA\%$ \uparrow $ & CER\%$\downarrow$ & NSPA\%$ \uparrow $ & CER\%$\downarrow$ & NSPA\%$ \uparrow $ & CER\%$\downarrow$ & NSPA\%$ \uparrow $ \\ 
        \midrule
        ESS & 17.38 & 60.4 & 16.61 & 95.4 & 16.77 & 98.9 & 17.26 & 98.9 \\
        MLS & 17.38 & 60.4 & 16.68 & 91.7 & 16.77 & 93.9 & 17.08 & 95.2 \\
        TSP & 17.38 & 60.4 & 16.59 & 93.4 & 16.64 & 98.1 & 16.79 & 97.6 \\           
        \bottomrule
    \vspace{-8pt}
    \end{tabular}
    }
    \label{tab:comparison}
\vspace{-20pt}
\end{table}
\subsection{Datasets}
\vspace{-4pt}
\textbf{Training set}: The training set includes clean speech, background noise, transient noises including claps, coughs, etc., and IRs used to simulate reverberation. 
The clean speech is from AISHELL-2~\cite{aishell2} and augmented to simulate scenarios with one to four passengers speaking simultaneously in a car. The background and transient noise are recorded in cars and last 10 hours. The IRs include both simulated and real-recorded IRs. The simulated IRs are generated using the image-source method\footnote{https://github.com/DavidDiazGuerra/gpuRIR/}, based on the actual size of the car cabins. We set the number of zones within the cabin to $Z = 4$ and simulate 25,000 IRs for each zone's microphone, 
resulting in a total of 100,000 simulated IRs. For generating the real-recorded IRs, we evaluate three different activation signals: exponential sine sweep signal (ESS)~\cite{ess}, maximum length sequence (MLS)~\cite{mls}, and time-streched pulses (TSP)~\cite{tsp}. We record 39 IRs in each zone or each activation signal, leading to 156 IRs across the four zones. 


\noindent\textbf{Test set}: Audios used for testing are recorded in real-world in-car scenarios consisting of two parts: the speech recognition test set and the zone positioning test set. The speech recognition test set lasts 7.4 hours, of which 37.5\% of the recordings are recorded during driving, with a lower signal-to-noise ratio (SNR). The zone positioning test set lasts 4.9 hours, with 57\% of the recordings made during car motion, 25\% of the recordings having the windows opened, and 75\% of the recordings involving ``non-standard postures". The remaining 43\% of the scenes are recorded when the car is stationary, with 67\% involving ``non-standard postures".

\vspace{-4pt}
\subsection{Data Augmentation}
\vspace{-4pt}
We use IRs and clean speech to simulate four-channel reverberant speech data, mimicking the recordings received by four microphones inside the cabin. In the 1st-stage of training, we convolve the clean speech with simulated IRs to simulate reverberant speech, while in the 2nd stage, we propose a data augmentation method that combines real-recorded IRs with simulated IRs. The channel corresponding to the speaker's sitting zone uses real-recorded IRs to simulate reverberant speech, while the other three channels use simulated IRs to simulate reverberant speech. We name this method ``mixed real-recorded IRs" in the experiment. Additionally, we compare two other methods for data simulation using real-recorded IRs. One method combines simulated IRs with real-recorded IRs in a certain ratio, where 25\% of the data has all channels augmented with real-recorded IRs, and the remaining 75\% uses simulated IRs. We call this method ``added real-recorded IRs". The other method only uses real-recorded IRs for reverberant speech simulation,  which is named ``only real-recorded IRs". After simulating the four-channel reverberant data, we add noise in an on-the-fly manner during training. The SNR range of background noise is [-20, 25] db, while the SNR range of transient noise is [-5, 5] db.
\vspace{-4pt}
\subsection{Experimental Setup}
\vspace{-4pt}
We design three models with different sizes, namely \texttt{CabinSep-S}, \texttt{CabinSep-M}, and \texttt{CabinSep-L}, with the number of full-sub modules $N$ being $1$, $2$, and $3$, respectively. For \texttt{CabinSep-S} and \texttt{CabinSep-M}, the channel compression ratio $d$ for $\textrm{Linear}_\textrm{A}$ and $\textrm{Linear}_\textrm{B}$ in the TAC is set to $4$, while in \texttt{CabinSep-L}, $d$ is set to $2$. In \texttt{CabinSep-S}, the conformer consists of $4$ layers, whereas the other two models have $2$ layers each. In all three models, the convolution layer's output dimension $H$ of the sub-band block is $16$, the feedforward dimension in the conformer is $\frac{H}{2}=8$, and the multi-head attention uses $4$ heads. We set the output dimension $C$ of the Conv2d before full-sub modules to $24$.

We adopt two other speech separation approaches, FasNet-TAC~\cite{fasnet-tac} and DualSep~\cite{dualsep}, as baselines for comparison. FasNet-TAC is a classic open source\footnote{https://github.com/yluo42/TAC/tree/master} end-to-end time-domain speech separation network, while DualSep is an in-car speech separation approach that reaches SOTA performance. We train and test these two models using the same datasets. For DualSep, we train the two models with different model sizes proposed in~\cite{dualsep}, namely DualSep-S and DualSep-L, for comparison. Since DualSep-L incorporates a non-causal IVA operation, making the model overall non-causal and unsuitable for practical applications, we substitute the non-causal IVA in DualSep-L with a causal IVA.

For both \texttt{CabinSep} and the DualSep systems, the input audio is transformed into the T-F domain via STFT. We use a hamming window with an FFT size of 512, window length of 32 ms, and hop length of 16 ms. 
During the training and finetuning, the Adam optimizer is adopted with an initial learning rate of 0.0001, which is halved every 20,000 steps.
The separated speech is sent to two different back-end ASR models, WeNet\footnote{https://github.com/wenet-e2e/wenet/blob/main/examples/wenetspeech/s0/} and SenseVoice\footnote{https://huggingface.co/FunAudioLLM/SenseVoiceSmall}, to evaluate the character error rate (CER), respectively.

\vspace{-4pt}
\subsection{Results}
\vspace{-4pt}


\textbf{Stage 1: Training on simulated IRs.}
Table 1 shows the CER results of the 1st-stage systems on the speech recognition test set, of which the upper part compares our proposed \texttt{CabinSep} systems with baselines, and the lower part shows the results of ablation experiments. We use two different ASR models, WeNet~\cite{wenet} and SenseVoice~\cite{sensevoice}, for speech recognition, comparing the static CER when the car is stationary, the motion CER when the vehicle is in motion, and the average CER on the whole test set. From the upper part of Table 1, we can see that \texttt{CabinSep-S} has significantly lower computational complexity than baselines while consistently achieving a lower CER. Specifically, \texttt{CabinSep-S}, with a computational complexity of only 0.4 GMACs and 0.21 RTF on a single-core Qualcomm SA8295P in-car CPU, achieves average CER relative reductions of 17.5\% and 14.2\% through WeNet and SenseVoice, respectively, compared with DualSep-L.
With the larger model size and computational complexity, \texttt{CabinSep-M} and \texttt{CabinSep-L} can receive a consistent improvement in CER performance. The best-performing model, \texttt{CabinSep-L}, achieved an average CER of 17.38\% with WeNet and 11.31\% with SenseVoice. Therefore, we selected \texttt{CabinSep-L} for the ablation experiments to evaluate the effectiveness of each of our proposed modules. 

\noindent\textbf{Ablation Results}: 
Analyzing the average CER results from WeNet in the lower part of Table 1, it is evident that each proposed module is practical. The most significant improvement is observed with the cascaded MVDR (7-1), reducing CER from 31.16\% to 17.38\% with only an additional 0.04 GMACs. Introducing a time skip operation (7-2) reduces computational complexity by 0.39 GMACs, with a minimal CER increase of 0.41\%. Replacing the conformer in the sub-band-conformer with LSTM (7-3) slightly decreases the parameter count and computational complexity by 0.01M and 0.03 GMACs, respectively, but increases CER by 0.56\%. Further omitting TAC modules (7-4) results in a 2.24\% increase in CER. Experiments 7-5 and 7-6 demonstrate that further removing the LPS and IPD encoders and estimating only the speech mask increases CER by 0.95\% and 0.93\%, respectively. Finally, adding chunks (7-7), limiting the conformer to look back at a maximum of 2 seconds of audio, only increases the CER by 0.09\%.

\noindent\textbf{Stage 2: Finetuning on real-recorded IRs.}
In the 2nd-stage experiments, we compare three types of activation signals to generate real-recorded IRs. For ``standard postures", all methods achieve 100\% positioning accuracy. As shown in Table 2, real-recorded IRs significantly improve the positioning accuracy in ``non-standard postures" (NSPA) scenarios, raising it from 60.4\% to over 90\%, with the highest accuracy reaching 98.9\%. We found that IRs generated using ESS and TSP activation signals perform similarly and better than those generated by MLS. The lowest CER is achieved using our proposed ``mixed real-recorded IRs" data augmentation method. 
Due to the limited amount of real-recorded IRs, \texttt{CabinSep-L} trained exclusively with real-recorded IRs for data augmentation performs worse in CER metric than using a combination of simulated and real-recorded IRs.

\vspace{-4pt}
\section{Conclusions}
\vspace{-4pt}

This paper proposes an in-car speech separation approach with excellent generalization ability, enhancing the back-end ASR models' performance. We validate the system using two different ASR models, WeNet and SenseVoice, without any joint training, demonstrating its plug-and-play capability and compatibility with various ASR systems. Trained with simulated data and tested on real-recorded test sets, the proposed \texttt{CabinSep-S} achieves a 17.5\% relative reduction in CER compared to the previous SOTA approach, DualSep-L, while requiring 1.12 GMACs less computational complexity. 
Additionally, we compared various data augmentation strategies that combine simulated and real-recorded IRs and addressed the issue of speaker mispositioning in-car scenarios, particularly for speakers in ``non-standard postures".

\bibliographystyle{IEEEtran}
\bibliography{mybib}

\end{document}